\documentclass[useAMS,usenatbib,usegraphicx]{mn2e}
\usepackage{bm}
\usepackage{times}

\title[Formation of micron-sized grains]
{Condition for the formation of micron-sized dust grains in dense molecular cloud cores}
\author[Hirashita and Li]{Hiroyuki Hirashita$^1$\thanks{E-mail:
    hirashita@asiaa.sinica.edu.tw}
and Zhi-Yun Li$^2$\\
$^1$Institute of Astronomy and Astrophysics, Academia Sinica,
P.O. Box 23-141, Taipei 10617, Taiwan\\
$^2$Astronomy Department, University of Virginia, Charlottesville, VA 22904,
USA
}
\date{2013 June 24}

\pagerange{\pageref{firstpage}--\pageref{lastpage}} \pubyear{2012}

\begin{document}
\label{firstpage}
\maketitle

\begin{abstract}
We investigate the condition for the formation of
micron-sized grains in dense cores of molecular 
clouds. This is motivated by the detection of the  
mid-infrared emission from deep inside a number of 
dense cores, the so-called `coreshine,' which is 
thought to come from scattering by micron
($\micron$)-sized  grains. Based on numerical
calculations of coagulation  starting from the typical grain
size distribution in  the diffuse interstellar medium, we obtain
a conservative lower 
limit to the time $t$ to form $\micron$-sized grains: 
$t/t_\mathrm{ff}>3 (5/S) (n_\mathrm{H}/10^5~\mathrm{cm}^{-3})^{-1/4}$
(where $t_\mathrm{ff}$ is the free-fall time at hydrogen
number density $n_\mathrm{H}$ in the core, and $S$
the enhancement factor 
to the grain-grain collision cross-section to account 
for {non-compact aggregates}). At the typical core density 
$n_\mathrm{H}=10^5~\mathrm{cm}^{-3}$, it takes at least
{a few}
free-fall times to form the $\micron$-sized 
grains responsible for coreshine. The implication is 
that those dense cores observed in coreshine are relatively 
long-lived entities in molecular clouds, rather than 
dynamically transient objects that last for one free-fall
time or less. 
\end{abstract}

\begin{keywords}
dust, extinction --- infrared: ISM --- ISM: clouds
--- ISM: evolution --- turbulence
\end{keywords}

\section{Introduction}
\label{intro}

Dense cores of molecular clouds are the basic units for the formation
of Sun-like, low-mass stars. A fundamental question about these cores 
that has not been answered conclusively is: are they long-lived
entities or simply transient objects that disappear in one free-fall 
time or less? 

The core lifetime is important to determine, because it affects the
rate of star formation as well as the time available for chemical 
reactions, which in turn affect the chemical structure of not only 
the cores themselves but also the disks (and perhaps even objects 
such as comets) that form out of them \citep{caselli12}.
It also has implications on how the cores are formed
\citep{ward07}. If the cores are relatively long lived, it would 
favor those formation scenarios that involve persistent support 
against gravity from, for example, magnetic fields
\citep{shu87,mouschovias99}, or long mass-accumulation 
time \citep[e.g.][]{gong11}. If the core lifetime turns out 
comparable to the free-fall time or less, then rapid formation 
and collapse, through for example turbulent compression, would be 
preferred \citep{maclow04}.

One way to constrain the core lifetime is to compare the number of 
starless cores to that of young stellar objects (YSOs), whose 
lifetimes  can be independently estimated
\citep{ward07,evans09}. \citet{ward07} found that 
cores of $10^4$--$10^5$~cm$^{-3}$ typically last for $\sim$ 2--5
free-fall times. Such estimates depend, however, on the lifetimes of YSOs, 
which are uncertain. Here, we explore another, completely independent, 
way of constraining the core lifetime, through the grain growth 
implied by the recently discovered phenomenon of `coreshine.' 

The so-called `coreshine' refers to the emission at mid-infrared
[especially the 3.6 $\micron$ \textit{Spitzer} Infrared Array
Camera (IRAC) band] from deep inside dense cores of molecular
clouds \citep{pagani10,steinacker10}. It is found in about half
of the cores where the emission is searched for \citep{pagani10}.  
The emission is thought to come from light scattered by
dust grains up to 1~$\micron$ in size. Such grains 
are much larger than those in the diffuse interstellar 
medium \citep*[e.g.][hereafter MRN]{mathis77}. Since it takes 
time for small MRN-type grains to grow to $\micron$-size, the 
observed coreshine should provide a constraint on the core 
lifetime. The goal of our investigation is to quantify this 
constraint. Specifically, we want to answer the question: how 
long does it take for the grains in a dense core to grow to  
$\micron$-size at a given density?

Grain growth through coagulation has been studied for a long time
\citep[e.g.][]{chokshi93,dominik97}. Even before the discovery of 
coreshine, \citet{ormel09} was able to demonstrate 
that coagulation can in principle produce $\micron$-sized grains
in dense cores, provided that the grains are coated by `sticky' 
materials such as water ice and that the cores are relatively 
long-lived \citep[see also][]{ormel11}.
In this paper, we aim to strengthen \citet{ormel09}'s 
results by deriving a robust lower limit to the lifetimes for 
those cores with $\micron$-sized grains inferred from coreshine 
through a simple framework that isolates the essential physics 
of coagulation. We find that cores of typical 
density $10^5$~cm$^{-3}$ must last for at least {a few}
free-fall times in order to produce $\micron$-sized grains. Our 
coagulation models are explained in Section~\ref{sec:model} 
and the results are described in Section~\ref{sec:results}. We 
discuss the robustness and implication of the results in
Section~\ref{sec:discussion}, and conclude in Section 
\ref{sec:conclusion}.

\section{Models}\label{sec:model}

\subsection{Coagulation}

We consider the time evolution of grain size distribution
by coagulation in a dense core. 
We adopt the formulation used in our previous paper,
\citet{hirashita12} (see also \citealt{hirashita09}),
with some changes to make
it suitable for our purpose.
{We briefly summarize the formulation here,
and refer to \citet{hirashita12} for further details}.

We assume that the grains are spherical with
a constant material density $\rho_\mathrm{gr}$.
%%dependent on the
%%grain species, so that the grain
%%mass $m$ and the grain radius $a$ are related as
%%\begin{eqnarray}
%%m=\frac{4}{3}\pi a^3\rho_\mathrm{gr}.\label{eq:mass}
%%\end{eqnarray}
We define the grain size distribution such that
$n(a,\, t)\,\mathrm{d}a$ is the number density of
grains whose radii are between $a$ and
$a+\mathrm{d}a$ at time $t$.
%%It is more useful
%%to use the number density of grains with mass
%%between $m$ and $m+\mathrm{d}m$,
%%$\tilde{n}(m,\, t)$, which is related to
%%$n(a,\, t)$ by
%%$\tilde{n}(m,\, t)\,\mathrm{d}m=n(a,\, t)\,\mathrm{d}a$;
%%that is, $\tilde{n}=n/(4\pi a^2 \rho_\mathrm{gr})$.
For numerical calculation, we consider $N=128$
discrete logarithmic bins for the grain radius (or mass),
and solve the discretized coagulation equation.
%%The mass density
%%of grains contained in the $i$th bin, $\tilde{\rho}_i$,
%%is defined as
%%\begin{eqnarray}
%%\tilde{\rho}_i \equiv m_i\tilde{n}(m_i)\Delta m_i,
%%\end{eqnarray}
%%where $\Delta m_i$ is the bin width, and $a_i$ and
%%$m_i=(4\pi /3)a_i^3\rho_\mathrm{gr}$ are the
%%representative grain radius and mass in the $i$th bin,
%%respectively. We adopt a constant time step, $\Delta t$,
%%which is shorter than the grain--grain collision
%%time-scale. The change of $\tilde{\rho}_i$ in a time step,
%%$\Delta\tilde{\rho}_i$, is written as
%%\begin{eqnarray}
%%\frac{\Delta\tilde{\rho}_i}{\Delta t}=
%%-m_i\tilde{\rho}_i
%%\sum_{k=1}^{N}\alpha_{ki}\tilde{\rho}_k+
%%\sum_{j=1}^{N}\sum_{k=1}^N\alpha_{kj}\tilde{\rho}_k
%%\tilde{\rho}_jm_\mathrm{coag}^{kj}(i)\, ,
%%\end{eqnarray}
%%\begin{eqnarray}
%%\alpha_{ki}=\left\{
%%\begin{array}{ll}
%%{\displaystyle \frac{\sigma_{ki}v_{ki}}{m_im_k}} &
%%\mbox{if $v_{ki}<v_\mathrm{coag}^{ki}$,} \\
%%0 & \mbox{otherwise.}
%%\end{array}
%%\right.
%%\end{eqnarray}
%%where $\sigma_{ki}$ and $v_{ki}$ are the
%%grain--grain collisional cross-section and
%%the relative collision speed between two grains
%%in the $k$th and $i$th bins.
%%Here, $m_\mathrm{coag}^{kj}(i)=m_k$\footnote{Note
%%that the subscript is $k$ (not $i$). We
%%count the same pair twice.} if
%%$m_k+m_j$ is in the mass range of the $i$th bin;
%%otherwise
%%$m_\mathrm{coag}^{kj}(i)=0$.
{In considering the grain--grain collision rate
between two grains with radii $a_1$ and $a_2$,}
we estimate the relative velocity by
\begin{eqnarray}
v_{12}=\sqrt{v(a_1)^2+v(a_2)^2-2v(a_1)v(a_2)\mu\,},
\label{eq:v12}
\end{eqnarray}
where the grain velocity as a function of grain radius,
$v(a)$, is given below in
equation (\ref{eq:v_turb}),
and $\mu\equiv\cos\theta$ ($\theta$ is an angle
between the two grain velocities) is randomly chosen
between $-1$ and 1 in each
time-step,\footnote{This treatment is different
from \citet{hirashita12}, who represented
the collisions by $\mu =-1$, 0, and 1. Such a
discrete treatment of $\mu$ cause
artificial spikes in the grain size distribution.}
and the cross-section by
\begin{eqnarray}
\sigma_{12}=S\pi (a_1+a_2)^2,\label{eq:cross}
\end{eqnarray}
where $S$ is the enhanced factor of cross-section,
which represents the increase of cross-section by
{non-compact aggregates}.
{Note that we always define the grain radius
$a$ and the grain material density $\rho_\mathrm{gr}$
for the compact geometry, even if $S>1$, to avoid the
extra uncertainty caused by the grain geometry
[see also the comment in the item (iii) in
Section \ref{subsec:initial}].
We adopt the turbulence-driven
grain velocity derived by
\citet{ormel09}, who assume that
the driving scales of turbulence is given by
the Jeans length and that the typical velocity
of the largest eddies ($\sim$ the Jeans length)
is given by the sound speed}
\citep[see also][]{hirashita12}:
\begin{eqnarray}
v(a) & = & 1.1\times 10^3\,\left(
\frac{T_\mathrm{gas}}{10~\mathrm{K}}\right)^{1/4}
\left(\frac{a}{0.1~\micron}\right)^{1/2}\nonumber\\
& \times & \left(
\frac{n_\mathrm{H}}{10^5~\mathrm{cm}^{-3}}\right)^{-1/4}
\left(\frac{\rho_\mathrm{gr}}{3.3~\mathrm{g~cm}^{-3}}\right)^{1/2}~
\mathrm{cm~s}^{-1},\label{eq:v_turb}
\end{eqnarray}
where $T_\mathrm{gas}$ is the gas temperature
assumed to be 10 K in this paper. Thermal
velocities are small enough to be neglected.
{The robustness of our conclusion in terms of the
grain velocity is further discussed in
Section \ref{subsec:threshold}.}

{
The form of equation (\ref{eq:v12}) suggests that the motions
of dust particles are random. This treatment is not valid
in general, since turbulent motions
are correlated. However, we do not include the full treatment
of the probability distribution function of the true relative
particle velocity in turbulence for the following three reasons:
(i) Our simple formulation
is sufficient to give a lower limit for the coagulation
time-scale (Section \ref{subsec:threshold}).
(ii) The probability
distribution function of the true relative particle velocity
in turbulence is unknown, and has only recently been
investigated 
\citep{hubbard13,pan13}.
(iii) In the environments of interest to this paper,
the forcing of turbulent eddies can be represented by a model
where the particle motions experience `random kicks':
in this so-called `intermediate regime' \citep{ormel07},
the prescription given by equation (\ref{eq:v12}) is
applicable. Indeed, we can confirm that the condition for
the intermediate regime is satisfied as follows.
The intermediate regime is defined by
$\mathrm{Re}^{-1/2}<\mathrm{St}<1$, where
Re is the Reynolds number and
St is the Stokes number \citep{ormel07}. This
condition is translated into
$11(a/1~\micron)^2(T_\mathrm{gas}/
10~\mathrm{K})^{-1}~\mathrm{cm}^{-3}
<n_\mathrm{H}<2.9\times 10^{12}(a/1~\micron)^4
(T_\mathrm{gas}/10~\mathrm{K})^{-1}~\mathrm{cm}^{-3}$.
Since we are interested in the range of
grain radius, $0.1~\micron\la a\la 1~\micron$, the
intermediate regime is applicable to
the density range considered in this paper.
}

We adopt the following coagulation threshold
velocity, $v_\mathrm{coag}^{ki}$, given by
\citep{chokshi93,dominik97,yan04}
\begin{eqnarray}
v_\mathrm{coag}^{ki}=21.4\left[
\frac{a_k^3+a_i^3}{(a_k+a_i)^3}\right]^{1/2}
\frac{\gamma^{5/6}}{\mathcal{E}^{\star 1/3}R_{ki}^{5/6}\rho_\mathrm{gr}^{1/2}}\, ,
\label{eq:vcoag}
\end{eqnarray}
where $\gamma$ is the surface energy per unit area,
$R_{ki}\equiv a_ka_i/(a_k+a_i)$ is the reduced radius of the
grains, $\mathcal{E}^\star$ is the the reduced elastic modulus.
This coagulation threshold is valid for collision
between two homogeneous spheres and would not be applicable
to collisions between {aggregates}.
{At low velocities, grains stick with each other and
develop a non-compact or fluffy aggregates. These aggregates
stick with each other at low relative velocities, and start to deform
or bounce as the relative velocities increases. Because the
deformation absorbs the collision energy, the aggregates
can stick with each other at a velocity larger than the
above coagulation threshold.
At very high velocities,
cratering and catastrophic destruction will halt the growth
\citep{paszun09,wada11,seizinger13}.}
In this paper,
we only limit the application of this threshold to
compact spherical grains [i.e.\ cases (i) and (ii) in
Section \ref{subsec:initial}; see \citet{ormel09}
and references therein for a detailed treatment of
coagulation of {aggregates}.]

\subsection{Initial condition and selection of parameters}
\label{subsec:initial}

For the initial grain size distribution, we adopt the
following power-law distribution, which is typical
in the diffuse ISM (MRN):
\begin{eqnarray}
n(a)={\cal C}a^{-3.5}~(a_\mathrm{min}\leq a\leq
a_\mathrm{max})\, ,
\end{eqnarray}
where ${\cal C}$ is the normalizing constant,
with $a_\mathrm{min}=0.001~\micron$
and $a_\mathrm{max}=0.25~\micron$.
The normalization factor ${\cal C}$ is determined
according to the mass density of the grains in the
ISM:
\begin{eqnarray}
{\cal D}\mu m_\mathrm{H}n_\mathrm{H}=
\int_{a_\mathrm{min}}^{a_\mathrm{max}}\frac{4\pi}{3}a^3
\rho_\mathrm{gr}{\cal C}a^{-3.5}\,\mathrm{d}a\, ,
\label{eq:norm_grain}
\end{eqnarray}
where $n_\mathrm{H}$ is the hydrogen number density,
$m_\mathrm{H}$ is the hydrogen atom
mass, $\mu$ is the atomic weight per hydrogen
(assumed to be 1.4) and ${\cal D}$ (0.01;
\citealt{ormel09}) is the dust-to-gas
mass ratio.

We adopt $n_\mathrm{H}=10^5$ cm$^{-2}$ for the
typical density of dense cores emitting coreshine
\citep{steinacker10}, but also survey a wide 
range in $n_\mathrm{H}$. We normalize
the time to the free-fall time, $t_\mathrm{ff}$:
\begin{eqnarray}
t_\mathrm{ff}=\sqrt\frac{3\pi}{32G\mu m_\mathrm{H}n_\mathrm{H}}
=1.38\times 10^5\left(\frac{n_\mathrm{H}}{10^5~\mathrm{cm}^{-3}}
\right)^{-1/2}~\mathrm{yr}.\label{eq:tff}
%%1.380
\end{eqnarray}

To isolate the key pieces of physics that determine the rate of
coagulation, we examine the following three models: 
% by changing
%the treatment of coagulation threshold
%velocity and grain cross section for coagulation:

\begin{enumerate}
\item[(i) {\bf Standard silicate model:}] We adopt
coagulation threshold given by equation (\ref{eq:vcoag})
with silicate material parameters
($\rho_\mathrm{gr}=3.3$ g cm$^{-3}$,
$\gamma =25$ erg cm$^{-2}$, and
$\mathcal{E}^\star =2.8\times 10^{11}$ dyn cm$^{-2}$;
\citealt{chokshi93}). We estimate the
cross-section by the compact spherical case (i.e.\
$S=1$ in equation \ref{eq:cross}). 

\item[(ii) {\bf Sticky coagulation model:}]
We do not apply the coagulation threshold; that is,
if grains collide with each other, they coagulate.
This is motivated by the fact that grains
coated by water ice have a large coagulation threshold
velocity \citep{ormel09}. We adopt $S=1$.

\item[(iii) {\bf Maximal coagulation model:}] As shown by
\citet{ormel09}, the volume filling factor of the grains 
after coagulation is $\sim 0.1$ because of
the {non-compact structure of aggregates}. Thus, we adopt
$S=5$ [$\sim (1/0.1)^{2/3}$]. Like the sticky coagulation
model, we do not apply the coagulation threshold.
This model provides a conservative estimate
for the coagulation time-scale (see
Section \ref{subsec:threshold} for discussion).
{Note that $a$ and $\rho_\mathrm{gr}$ are
defined for the compact grains. In fact, the
grain velocity (equation \ref{eq:v_turb}) also
has a dependence on the {volume filling
factor of aggregates} through
$a$ and $\rho_\mathrm{gr}$ in such a way that
{the non-compact structure} enhances the gas--grain coupling, leading
to a lower velocity. Thus, the maximal coagulation
model overestimates the grain velocity
(i.e.\ the coagulation rate),
which strengthens the case for the model being `maximal'.}
\end{enumerate}

\begin{figure*}
\includegraphics[width=0.33\textwidth]{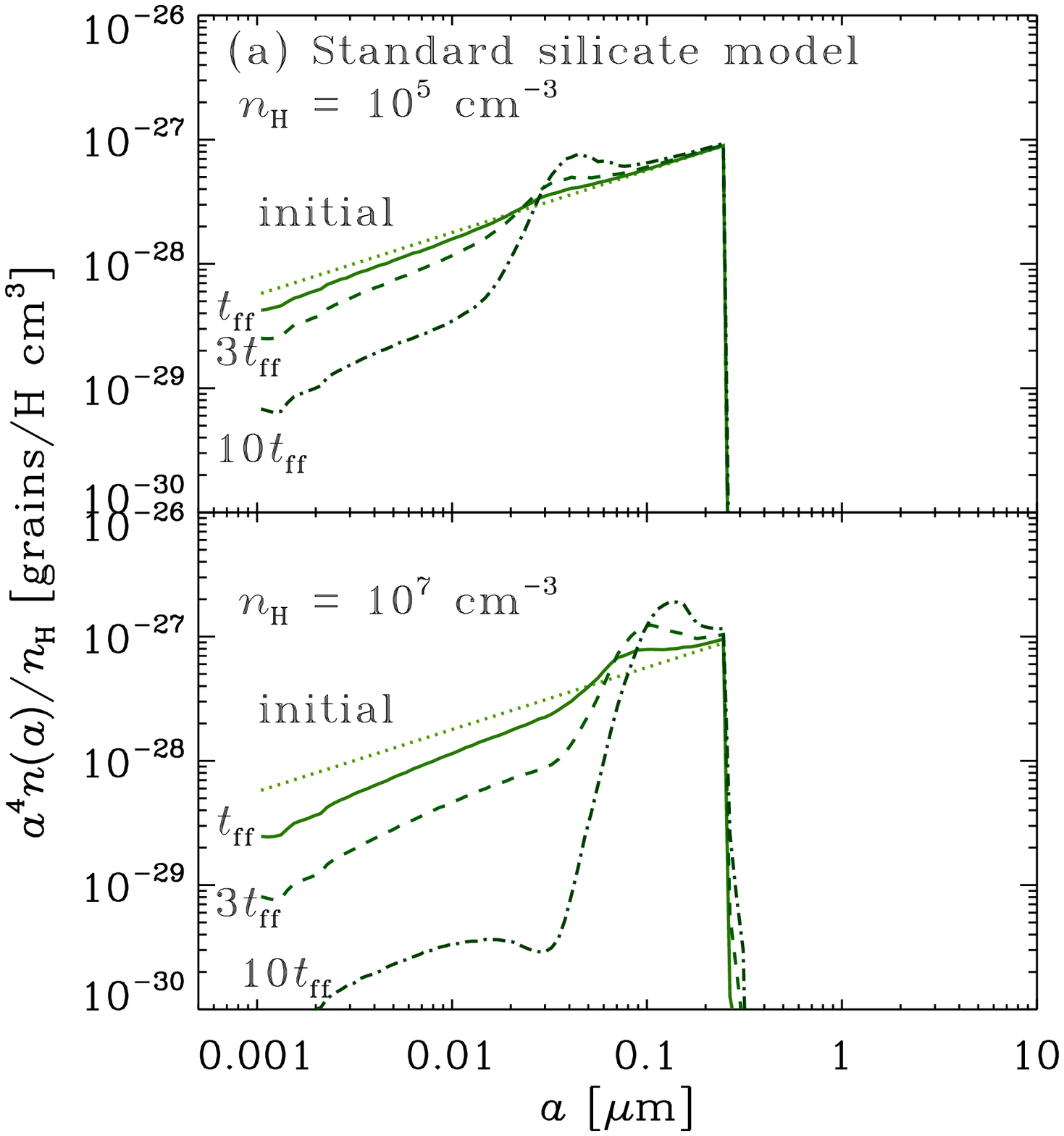}
\includegraphics[width=0.33\textwidth]{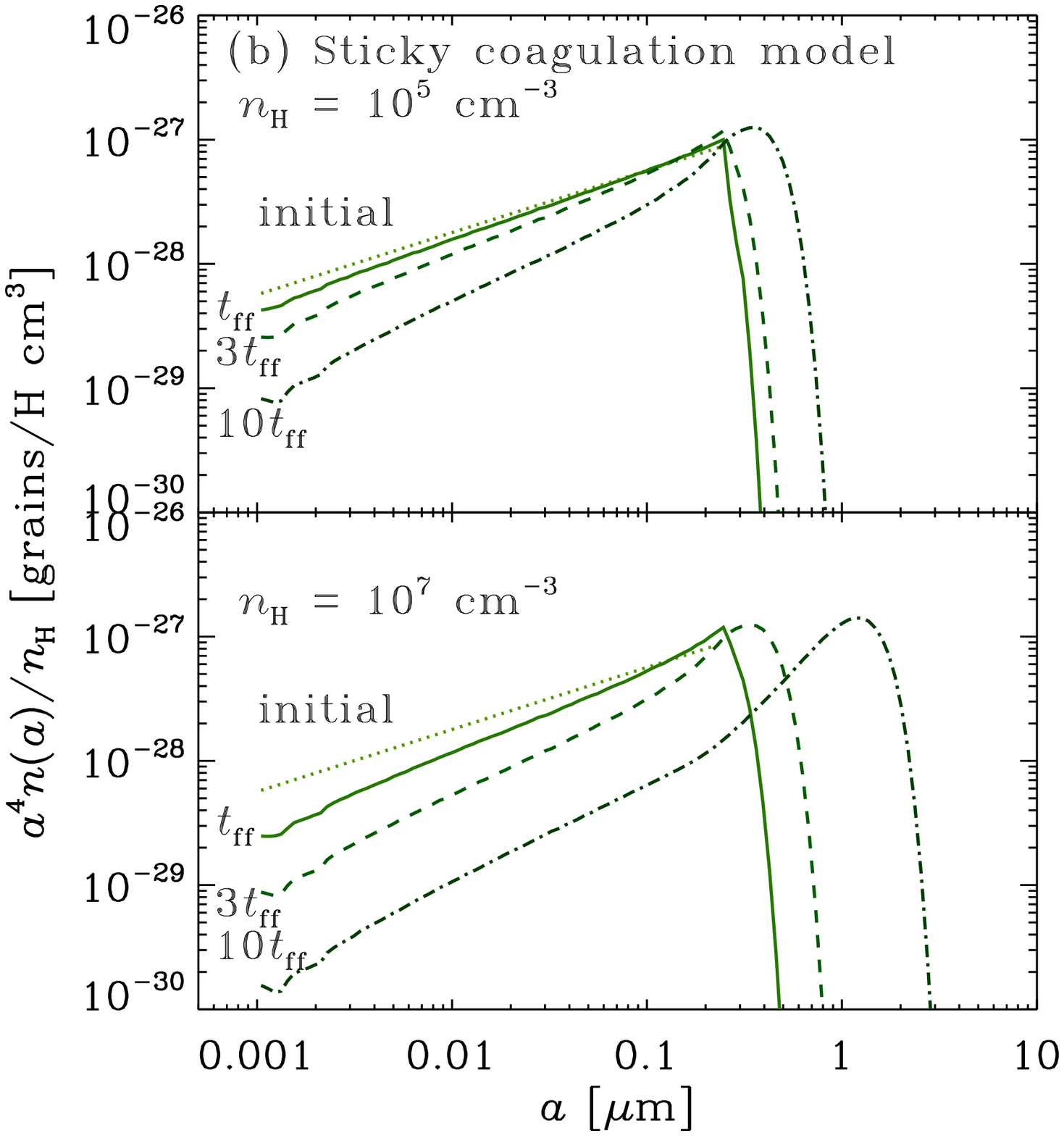}
\includegraphics[width=0.33\textwidth]{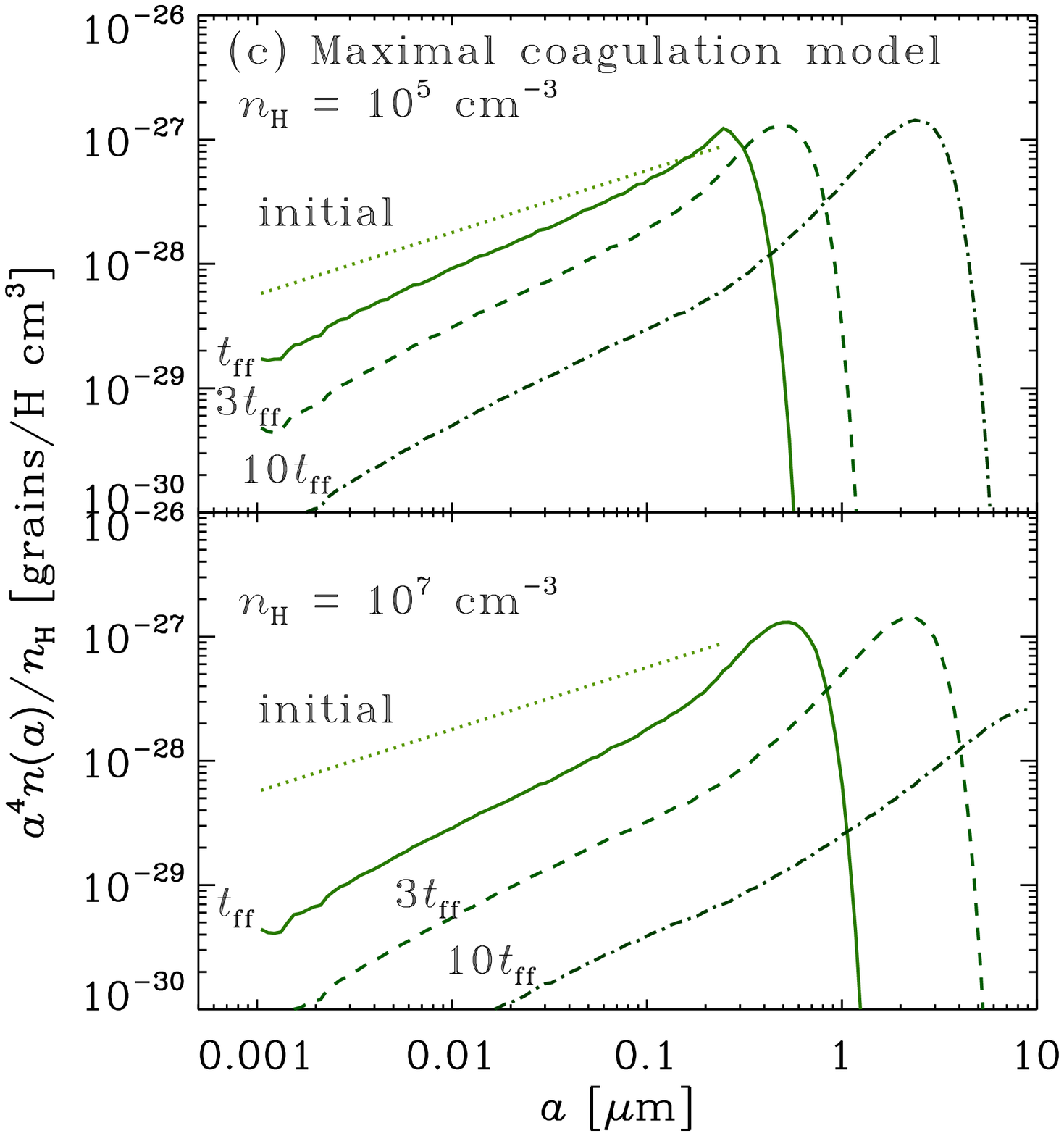}
\caption{Evolution of grain size distribution.
The solid, dashed, and dot-dashed, lines show the grain
size distributions at $t= 1 t_\mathrm{ff}$, $3 t_\mathrm{ff}$,
and $10 t_\mathrm{ff}$, respectively,
for (a) the standard silicate model,
(b) the sticky coagulation model, and (c) the maximal
coagulation model. The dotted
line presents the initial condition.
The upper and
lower panels show the cases with
$n_\mathrm{H}=10^5$ cm$^{-3}$ and $10^7$ cm$^{-3}$,
respectively.
\label{fig:size}}
\end{figure*}

\section{Results}\label{sec:results}

\subsection{Evolution of grain size distribution}

We present the evolution of grain size distribution
for $n_\mathrm{H}=10^5$ and $10^7$ cm$^{-3}$
at $t= 1t_\mathrm{ff}$, $3t_\mathrm{ff}$, and
$10t_\mathrm{ff}$. The results are shown in
Fig.\ \ref{fig:size} for all three models: 
(i) the standard silicate model,
(ii) the sticky coagulation model, and
(iii) the maximal coagulation model.
In order to show the grain mass distribution per
logarithmic radius, we show $a^4n(a)$.

In the standard silicate model shown in
Fig.\ \ref{fig:size}a, the grain growth stops at
$a\sim 0.1~\micron$ because of the coagulation
threshold: the grain velocities are too large
for coagulation if $a\ga 0.1~\micron$. Thus,
bare silicate cannot grow to $\micron$ sizes, a 
result found previously by \citet{ormel09}. 
We conclude that bare silicate cannot be the
source of coreshine.

Indeed, water ice has higher coagulation threshold,
so if grains are coated by water ice, coagulation
proceeds further \citep{ormel09,ormel11}. Motivated by
this, we examine the sticky coagulation model,
in which there is no coagulation threshold.
(The coagulation threshold of water ice is separately
discussed in Section \ref{subsec:threshold} to
minimize the uncertainty in the material properties
adopted.) Fig.~\ref{fig:size}b shows that
grains grow beyond 0.1 $\micron$. For
$n_\mathrm{H}=10^7$~cm$^{-3}$, $\micron$-sized
grains form at $10t_\mathrm{ff}$, while for
the standard density $n_\mathrm{H}=10^5$~cm$^{-3}$,
the typical grain radius does not reach 1 $\micron$
even at $10t_\mathrm{ff}$.

In reality, {aggregates} are thought to form as a
result of coagulation \citep{ossenkopf93}. Thus,
the cross-section is
effectively increased compared with the spherical
and compact case. Fig.\ \ref{fig:size}c shows
that the maximal coagulation model in which the
cross-section is elevated by a factor of 5
(i.e.\ $S=5$) successfully
produces $\micron$-sized grains within $10 t_\mathrm{ff}$
even for $n_\mathrm{H}=10^5$ cm$^{-3}$.
It remains difficult, however, to produce $\micron$-sized grains
in $3 t_\mathrm{ff}$ for $n_\mathrm{H}=10^5$~cm$^{-3}$
and in $1 t_\mathrm{ff}$ for $n_\mathrm{H}=10^7$ cm$^{-3}$.

\subsection{Condition for the formation of $\micron$-sized
grains}
\label{subsec:condition}

As mentioned in Introduction, the aim of this paper is to determine 
the condition for the formation of $\micron$-sized grains thought to 
be responsible for the observed coreshine \citep{pagani10,steinacker10}.
{According to \citet{steinacker10}, scattering dominates over
absorption by an order of magnitude at $\lambda =3.6~\micron$ if
$a\ga 1~\micron$.}
Since the peak of the grain size distribution in $a^4n(a)$ is well 
defined (see Fig.\ \ref{fig:size}), we simply find the condition 
for the radius at the peak, $a_\mathrm{peak}$, reaches or exceeds
1 $\micron$. 
{
We also examine a more conservative criterion by using
$a_\mathrm{peak}=0.5~\micron$ instead of 1 $\micron$, motivated in
part by the fact that $a\sim 0.5~\micron$ is the grain radius
at which
scattering is comparable to absorption at $\lambda =3.6~\micron$
\citep{steinacker10}.}

\begin{figure}
\includegraphics[width=0.45\textwidth]{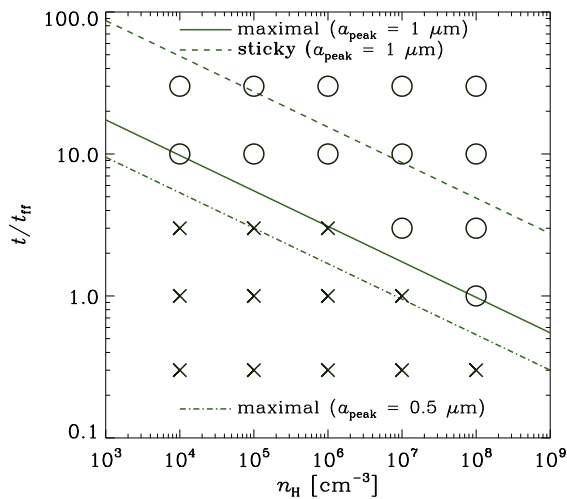}
\caption{
The condition for the formation of $\micron$-sized
grains. The success and failure of the
formation of $a>1~\micron$ grains in the maximal
coagulation model are shown by
`o' and `x', respectively. The solid and dashed lines
show the boundary of those two cases in the
maximal coagulation model and the sticky coagulation
model, respectively, {if we adopt
$a_\mathrm{peak}=1~\micron$ for the criterion for coreshine.
The dot-dashed line marks the boundary for the
maximal coagulation model for a more conservative
criterion: $a_\mathrm{peak}=0.5~\micron$.}
\label{fig:condition}}
\end{figure}

We will concentrate on the maximal coagulation model with an
enhancement factor for cross-section $S=5$; the result from the 
sticky coagulation model with $S=1$ can be obtained through a 
simple scaling. 
%which provides 
%the most conservative condition on the core lifetime for producing 
%$\micron$-sized grains. 
In Fig.\ \ref{fig:condition}, we show a grid
of models with different core densities and times (in units of the 
free-fall time at the core density). The solid line marks roughly 
the critical time $t_\mathrm{grow}$ at a given density $n_\mathrm{H}$ 
above which $\micron$-sized grains are produced. It is given by 
\begin{eqnarray}
\frac{t_\mathrm{grow}}{t_\mathrm{ff}}=A\left(\frac{5}{S}\right)
\left(\frac{n_\mathrm{H}}{10^5~\mathrm{cm}^{-3}}\right)^{-1/4},
\label{eq:condition}
\end{eqnarray}
{where $A=5.5$ and 3.0, respectively, if we adopt $a_\mathrm{peak}=1~\micron$
and $0.5~\micron$ for the criterion of micron-sized grain
formation.}
The condition for forming $\micron$-sized grains is therefore 
$t > t_\mathrm{grow}$. The same condition applies to the sticky 
coagulation model (with $S=1$) as well, since coagulation time 
is inversely proportional to the cross-section for grain-grain
collision. 

%
% the critical time $t_\mathrm{grow}$ 
% becomes 5 times longer, because it is inversely proportional 
% to the enhancement factor for cross-section
% (i.e.\ $t_\mathrm{grow}\propto S^{-1}$). 
%

Equation (\ref{eq:condition}) can be understood in
the following way. Since coagulation is a collisional
process, $t_\mathrm{grow}$ should be given by the collision
time-scale, $t_\mathrm{coll}=(vS\pi a^2n_\mathrm{dust})^{-1}
=4a\rho_\mathrm{gr}/(3\mathcal{D}\mu m_\mathrm{H}n_\mathrm{H}vS)$,
where $v$ and $n_\mathrm{dust}$ are the velocity and
the number density of grains, respectively
\citep{ormel09}. The growth time-scale in terms
of grain radius is $t_\mathrm{grow}\simeq 3t_\mathrm{coll}$
(note that $t_\mathrm{coll}$ is the time-scale of grain volume
being doubled by coagulation).
Then, $t_\mathrm{coll}/t_\mathrm{ff}$ is evaluated
by using equations~(\ref{eq:v_turb}) and (\ref{eq:tff}) as
$t_\mathrm{grow}/t_\mathrm{ff}\simeq 7.4(a/1~\micron)^{1/2}
(n_\mathrm{H}/10^5~\mathrm{cm}^{-3})^{-1/4}
(S/5)^{-1}
(T_\mathrm{gas}/10~\mathrm{K})^{-1/4} \cdot
(\rho_\mathrm{gr}/3.3~\mathrm{g~cm}^{-3})^{1/2}$;
that is, $A=7.4$ (5.2) for $a=1~\micron$ ($0.5~\micron$),
in a fair agreement with
the above numerical estimate. Thus, $t_\mathrm{grow}$ can 
be understood in terms of collision time-scale, which 
strengthens our numerical results.

Note that, to form $\micron$-sized grains in one free-fall
time, the density $n_\mathrm{H}$ must be of order $10^8$ 
cm$^{-3}$ or higher, even in the maximal coagulation 
model. In the sticky coagulation model, the required 
density would be higher still. Such densities are much 
higher than the typical core value (of order $10^5$~cm$^{-3}$). 
At $10^5$~cm$^{-3}$, Fig.~\ref{fig:condition} 
and equation~(\ref{eq:condition}) indicate that, under
reasonable conditions, it takes at 
least several free-fall times for the grains to grow to 
$\micron$-size (see Section~\ref{subsec:LongLivedCore} for more 
discussion). The implication is that those dense cores 
detected in coreshine should be rather long-lived entities 
rather than transient objects that disappear in one 
free-fall time; the latter objects would simply not have 
enough time to form the $\micron$-sized grains responsible 
for coreshine.

\section{Discussion}\label{sec:discussion}

\subsection{A lower limit to $\micron$-sized grain formation time} 
\label{subsec:threshold}

One may argue that coagulation would be faster if the grains 
were to collide at higher speeds than adopted in our model. 
However, it will be difficult for this to happen because of the 
existence of a coagulation threshold. As mentioned earlier, 
bare silicate grains already acquire velocities larger than 
the threshold at a rather small size $a\sim 0.1~\micron$; they 
do not grow beyond $0.1~\micron$ under reasonable conditions. 
To grow to larger sizes, the grains must be `more sticky' 
than silicate, as is the case when the grains are coated with 
water ice \citep{ormel09}. For such coated grains, we can 
estimate the coagulation threshold for equal-sized grains 
from equation (\ref{eq:vcoag}) using $\rho_\mathrm{gr}=3.3$ 
g cm$^{-3}$, $\gamma =370$ erg cm$^{-2}$, $\mathcal{E}^\star 
=3.7\times 10^{10}$ dyn cm$^{-2}$. The result is $v_\mathrm{coag} 
=9.4\times 10^2(a/1~\micron )^{-5/6}$ cm s$^{-1}$. For the 
micron-sized grains that we aim to form, this threshold is 
already smaller than the typical grain velocity $v\sim 3.5 
\times 10^3 (a/1~\micron)^{1/2}$ that was used in our model. 
In other words, our model is already generous with the 
grain-grain collision speed. (Collisions at the relatively high
  speed that we adopted may lead to the compaction of
{aggregates}, which should  
reduce the enhancement factor $S$ for grain-grain collision
cross-section and hence the rate of grain growth.).
Increasing {the collision speed} further should not lead to faster growth to
$\micron$-size.
 For this reason, we believe that the critical 
time $t_\mathrm{grow}$ for the formation of $\micron$-sized grains 
estimated in equation~(\ref{eq:condition}) is a robust lower 
limit. 

%
%In particular, it would be difficult to reduce 
%$t_\mathrm{grow}$ by a factor of 5 (so that $\micron$-sized 
%grains can form at the typical core density $10^5$~cm$^{-3}$ in 
%one free-fall time) by increasing the collision velocity by a 
%factor of 5 (and the kinetic energy by a factor of 25).
%

\subsection{The case for long-lived dense cores}
\label{subsec:LongLivedCore}

%
% long formation time at constant density first, tying to core
% lifetime second...evolution....fast formation in shocks?? 
% unlikely? 
%
Equation~(\ref{eq:condition}) indicates that it takes more than $\sim 5$
free-fall times to form $1~\micron$-sized grains at the typical core 
density $n_\mathrm{H}=10^5$~cm$^{-3}$ if the enhancement factor for 
cross-section is $S=5$. If the enhancement factor is larger, 
the coagulation would be faster. In particular, if $S=25$, the 
formation of micron-sized grains may occur in a single, rather than 
5, free-fall time. However, $S=25$ requires the grain volume filling 
factor to be $25^{-3/2}\sim 1$ per cent, which is extreme. For 
example, to form such a grain of $a=1~\micron$ with compact 
spherical grains with $a=0.1~\micron$, one need to connect 1,000 
grains \textit{linearly}, which is unlikely. We doubt that there is 
much room to increase $S$ well beyond $5$, which corresponds
{aggregates} of rather low volume filling factor ($\sim 0.1$) already. 
If the cross-section enhancement factor $S$ is not much larger 
than 5, it would take several free-fall times (or more) to form 
$\micron$-sized grains at typical core densities. The long 
formation time would indicate that those dense cores with observed 
coreshine are relatively long-lived entities, rather than transient 
objects that form and disappear in one free-fall time. {This
estimate of core lifetime based on grain growth is consistent with 
that inferred from the number of starless cores (relative to YSOs)
\citep{ward07}. It is also consistent with the observational results 
that only a small fraction of dense cores show any detectable sign 
of gravitational collapse and that even those collapsing cores 
tend to have infall speeds less than half the sound speed
\citep{difrancesco07}. Such slowly-evolving, relatively long-lived 
cores can form, for example, as a result of ambipolar 
diffusion in magnetically supported clouds
\citep{shu87,mouschovias99}, even in the presence of a strong,
supersonic turbulence \citep{nakamura05}. They are less compatible
with transient cores that are formed rapidly through fast compression by 
supersonic turbulence without any magnetic cushion \citep{maclow04}, 
unless the core material is slowly accumulated in the post-shock 
region over several free-fall times \citep[e.g.][]{gong11}. }
%
% what about argument of pre-growth before core?
% Unless pre-growth occurs in a denser region, the same
% discussion applies. We cannot deny that large grains
% are supplied from very dense regions through some outflow.

\subsection{Source of large grains}

Large grains ($a\ga 0.1~\micron$), once they are
injected into the diffuse ISM, are rapidly
shattered into smaller grains
\citep{hirashita09,asano13}. Thus, there should
be a continuous supplying mechanism of large grains
\citep{hirashita13}.
If dense molecular cores has lifetimes long enough
to produce $\micron$-sized grains, they can be
an important source of large grains. Including
the supply of large grains from dense cores will
be an interesting topic in modeling the
evolution of dust in galaxies.

%
% future work: let the referee force us to discuss drawbacks and
% future work here? Couple grain growth model to a dynamic model
%
\section{Conclusion}\label{sec:conclusion}

Motivated by recent coreshine observations, we have examined the condition 
for the formation of $\micron$-sized grains by coagulation in 
dense molecular cloud cores. We obtained a simple,
{conservative} lower limit 
to the core lifetime $t$ for the formation of  0.5 $\micron$-sized grains:
$t/t_\mathrm{ff}>3 (5/S)
(n_\mathrm{H}/10^5~\mathrm{cm}^{-3})^{-1/4}$, 
where $t_\mathrm{ff}$ is the free-fall time at the core density 
$n_\mathrm{H}$ and $S$ the enhancement factor for grain-grain 
collision that accounts for {aggregates}. The formation time for 
1 $\micron$-sized grains is roughly a factor of 2 longer. Since $S$ is unlikely 
much larger than 5, we conclude that dense cores of typical density 
$n_\mathrm{H}=10^5~\mathrm{cm}^{-3}$ must last for at least several 
free-fall times in order to produce the $\micron$-sized grains 
thought to be responsible for the observed coreshine. Such cores 
are therefore relatively long-lived entities in molecular clouds,
rather than dynamically transient objects.

\section*{Acknowledgments}

{We are grateful to C. W. Ormel for comments that greatly improved
the presentation of the paper.}
This research is supported through NSC grant 99-2112-M-001-006-MY3 and 
NASA grant NNX10AH30G.

%%\appendix

%%\section{Data for the radio--FIR relation}\label{app:radio_FIR}

\bsp

\label{lastpage}


\begin{thebibliography}{}
\bibitem[\protect\citeauthoryear{Asano et al.}{2013}]{asano13}
    Asano, R., Takeuchi, T. T., Hirashita, H., \& Nozawa, T. 2013,
    MNRAS, 432, 637
\bibitem[\protect\citeauthoryear{Caselli \& Ceccarelli}{2012}]{caselli12}
    Caselli, P. \& Ceccarelli,  C. 2012, A\&AR, 20, 56
%%\bibitem[\protect\citeauthoryear{Cardelli, Clayton, \& Mathis}{Cardelli et al.}{1989}]{cardelli89}
%%    Cardelli J. A., Clayton G. C., Mathis J. S., 1989, ApJ, 345, 245
\bibitem[\protect\citeauthoryear{Chokshi, Tielens, \& Hollenbach}{1993}]{chokshi93}
    Chokshi, A., Tielens, A. G. G. M., \& Hollenbach, D. 1993, ApJ,
    407, 806
%%\bibitem[\protect\citeauthoryear{Cox}{2000}]{cox00}
%%    Cox, A. N. 2000, Allen's Astrophysical Quantities, 4th ed., Springer,
%%    New York
\bibitem[\protect\citeauthoryear{Di Francesco et al.}{2007}]{difrancesco07}
    Di Francesco, J., Evans, N. J. II, Caselli, P. et al. 2007, in Reipurth B., Jewitt D., Keil K.,
    eds, Protostars and
    Planets V, University of Arizona Press, Tuscon, p.\ 17
\bibitem[\protect\citeauthoryear{Dominik \& Tielens}{1997}]{dominik97}
    Dominik, C., \& Tielens, A. G. G. M. 1997, ApJ, 480, 647
\bibitem[\protect\citeauthoryear{Draine}{1985}]{draine85}
    Draine, B. T. 1985, in Black D. C., Matthews M. S., eds,
    Protostars and Planets II. University of Arizona Press, Tucson, p.\ 621
%%\bibitem[\protect\citeauthoryear{Draine}{2009}]{draine09}
%%    Draine B. T., 2009, in Henning Th., Gr\"{u}n E., Steinacker J., eds,
%%    ASP Conf.\ Ser.\ 414, Cosmic Dust -- Near and Far.
%%    Astron.\ Soc.\ Pac., San Francisco, p.\ 453
%%\bibitem[\protect\citeauthoryear{Draine \& Anderson}{1985}]{draine85}
%%    Draine, B. T., \& Anderson, N. 1985, ApJ, 292, 494
%%\bibitem[\protect\citeauthoryear{Draine \& Lee}{1984}]{draine84}
%%    Draine, B. T., \& Lee, H. M. 1984, ApJ, 285, 89
%%\bibitem[\protect\citeauthoryear{Dwek}{1998}]{dwek98}
%%    Dwek E., 1998, ApJ, 501, 643
\bibitem[\protect\citeauthoryear{Evans  et al.}{2009}]{evans09}
    Evans, N. J., Dunham, M. M., et al. 2009, ApJS, 181, 321
\bibitem[\protect\citeauthoryear{Gong \& Ostriker}{2011}]{gong11}
    Gong, H. \& Ostriker, E. 2011, ApJ, 729, 120
%%\bibitem[\protect\citeauthoryear{Hirashita}{2000}]{hirashita00}
%%    Hirashita H., 2000, PASJ, 52, 585
\bibitem[\protect\citeauthoryear{Hirashita}{2012}]{hirashita12}
    Hirashita, H. 2012, MNRAS, 422, 1263
%%\bibitem[\protect\citeauthoryear{Hirashita \& Ferrara}{2002}]{hirashita02}
%%    Hirashita, H., \& Ferrara, A. 2002, MNRAS, 337, 921
%%\bibitem[\protect\citeauthoryear{Hirashita \& Hunt}{2004}]{hirashita04}
%%    Hirashita, H., \& Hunt, L. K. 2004, A\&A, 421, 555
%%\bibitem[\protect\citeauthoryear{Hirashita \& Kuo}{2011}]{hirashita_kuo}
%%    Hirashita, H., \& Kuo, T.-M. 2011, MNRAS, 416, 1340
\bibitem[\protect\citeauthoryear{Hirashita \& Nozawa}{2013}]{hirashita13}
    Hirashita, H. \& Nozawa, T. 2013, Earth Planets Space, 65, 183
\bibitem[\protect\citeauthoryear{Hirashita \& Yan}{2009}]{hirashita09}
    Hirashita, H., \& Yan, H. 2009, MNRAS, 394, 1061
\bibitem[Hubbard(2013)]{hubbard13} Hubbard, A. 2013, MNRAS, 432,
    1274
%%\bibitem[\protect\citeauthoryear{Jones \& Nuth}{2011}]{jones11}
%%    Jones, A. P., \& Nuth, J. A., III 2011, A\&A, 530, A44
%%\bibitem[\protect\citeauthoryear{Jones, Tielens, \& Hollenbach}{Jones et al.}{1996}]{jones96}
%%    Jones, A. P., Tielens, A. G. G. M., \& Hollenbach, D. J. 1996, ApJ,
%%    469, 740
%%\bibitem[\protect\citeauthoryear{Jones et al.}{1994}]{jones94}
%%    Jones, A. P., Tielens, A. G. G. M., Hollenbach, D. J., \&
%%    McKee, C. F. 1994, ApJ, 433, 797
%%\bibitem[\protect\citeauthoryear{Lisenfeld \& Ferrara}{1998}]{lisenfeld98}
%%    Lisenfeld, U., \& Ferrara, A. 1998, ApJ, 496, 145
\bibitem[\protect\citeauthoryear{Mathis, Rumpl, \& Nordsieck}{Mathis et al.}{1977}]{mathis77}
    Mathis, J. S., Rumpl, W., \& Nordsieck, K. H. 1977, ApJ, 217, 425
    (MRN)
\bibitem[\protect\citeauthoryear{Mac Low \& Klessen}{2004}]{maclow04}
    Mac Low, M. M. \& Klessen, R. S. 2004, RvMP, 76, 125
%%\bibitem[\protect\citeauthoryear{Mattsson}{2011}]{mattsson11}
%%    Mattsson, L. 2011, MNRAS, 414, 781
%%\bibitem[\protect\citeauthoryear{McKee}{1989}]{mckee89}
%%    McKee, C. F. 1989, in Allamandola L. J. \& Tielens A. G. G. M. eds.,
%%    IAU Symp.\ 135, Interstellar Dust, Kluwer, Dordrecht, 431
\bibitem[\protect\citeauthoryear{Mouschovias \& Ciolek}{1999}]{mouschovias99}
    Mouschovias, T. \& Ciolek, G. E. 1999, in Lada C. J., Kylafis N.,
    eds, The Origin of Stars
    and Planetary Systems, Kluwer, Dordrecht, p.\ 305
\bibitem[\protect\citeauthoryear{Nakamura \& Li}{2005}]{nakamura05}
  Nakamura, F. \& Li, Z.-Y. 2005, ApJ, 631, 411
%%\bibitem[Nozawa et al.(2007)]{nozawa07} Nozawa, T., Kozasa, T.,
%%    Habe, A., Dwek, E., Umeda, H., Tominaga, N., Maeda, K., \&
%%    Nomoto, K. 2007, ApJ, 666, 955
\bibitem[\protect\citeauthoryear{Ormel \& Cuzzi}{2007}]{ormel07}
    Ormel, C. W., \& Cuzzi, J. N. 2007, A\&A, 466, 413
\bibitem[\protect\citeauthoryear{Ormel et al.}{2011}]{ormel11}
    Ormel, C. W., Min, M., Tielens, A. G. G. M., Dominik, C., \&
    Paszun, D. 2011, A\&A, 532, A43
\bibitem[\protect\citeauthoryear{Ormel et al.}{2009}]{ormel09}
    Ormel, C. W., Paszun, D., Dominik, C., \& Tielens, A. G. G. M. 2009,
    A\&A, 502, 845
\bibitem[\protect\citeauthoryear{Ossenkopf}{1993}]{ossenkopf93}
    Ossenkopf, V. 1993, A\&A, 280, 617
\bibitem[\protect\citeauthoryear{Pagani et al.}{2010}]{pagani10}
    Pagani, L., Steinacker, J., Bacmann, A., Stutz, A.,
    \& Henning, T. 2010, Science, 329, 1622
\bibitem[\protect\citeauthoryear{Pan \& Padoan}{2013}]{pan13}
    Pan, L., \& Padoan, P. 2013, ApJ, submitted (arXiv:1305.0307) 
\bibitem[\protect\citeauthoryear{Paszun \& Dominik}{2009}]{paszun09}
    Paszun, D., \& Dominik, C. 2009, A\&A, 507, 1023
\bibitem[\protect\citeauthoryear{Seizinger \& Kley}{2013}]{seizinger13}
    Seizinger, A., \& Kley, W. 2013, A\&A, 551, A65
\bibitem[\protect\citeauthoryear{Shu et al.}{1987}]{shu87}
    Shu, F. H, Adams, F. \&  Lizano, S. 1987, ARA\&A, 25, 23
%%\bibitem[Rybicki \& Lightman(1979)]{rybicki79} Rybicki, G. B., \&
%%    Lightman, A. P. 1979, Radiative Processes in Astrophysics
%%    (New York: Wiley)
%%\bibitem[\protect\citeauthoryear{Spitzer}{1978}]{spitzer78}
%%    Spitzer, L. 1978, Physical
%%    Processes in the Interstellar Medium (New York: Wiley)
\bibitem[\protect\citeauthoryear{Steinacker et al.}{2010}]{steinacker10}
    Steinacker, J., Pagani, L., Bacmann, L., \& Guieu, S. 2010,
    A\&A, 511, A9
%%\bibitem[\protect\citeauthoryear{Takeuchi et al.}{2005}]{takeuchi05}
%%    Takeuchi, T. T.,
%%    Ishii, T. T., Nozawa, T., Kozasa, T., \& Hirashita, H. 2005,
%%    MNRAS, 362, 592
%%\bibitem[\protect\citeauthoryear{Valiante et al.}{2011}]{valiante11}
%%    Valiante, R., Schneider, R., Salvadori, S., \& Bianchi, S.
%%    2011, MNRAS, 416, 1916
\bibitem[\protect\citeauthoryear{Wada et al.}{2011}]{wada11}
    Wada, K., Tanaka, H., Suyama, T., Kimura, H., \& Yamamoto, T. 2011,
    ApJ, 737, 36
\bibitem[\protect\citeauthoryear{Ward-Thompson et al.}{2007}]{ward07}
    Ward-Thompson, D., Andr\'{e}, P., Crutcher, R., Johnstone, D.,
    Onishi, T., \& Wilson, C. 2007, in Reipurth B., Jewitt D., Keil K.,
    eds, Protostars and
    Planets V, University of Arizona Press, Tuscon, p.\ 33
%%\bibitem[\protect\citeauthoryear{Weingartner \& Draine}{1999}]{weingartner99}
%%    Weingartner, J. C., \& Draine, B. T. 1999, ApJ, 517, 292
%%\bibitem[\protect\citeauthoryear{Weingartner \& Draine}{2001}]{weingartner01}
%%    Weingartner, J. C., \& Draine, B. T. 2001, ApJ, 548, 296
%%\bibitem[\protect\citeauthoryear{Yamasawa et al.}{2011}]{yamasawa11}
%%    Yamasawa, D., Habe, A., Kozasa, T., Nozawa, T., Hirashita, H.,
%%    Umeda, H., \& Nomoto, K. 2011, ApJ, 735, 44
\bibitem[\protect\citeauthoryear{Yan, Lazarian \& Draine}{2004}]{yan04}
    Yan H., Lazarian A., Draine B. T., 2004, ApJ, 616, 895
%%\bibitem[\protect\citeauthoryear{Yasuda \& Kozasa}{2012}]{yasuda12}
%%    Yasuda, Y., \& Kozasa, T. 2012, ApJ, 745, 159
\end{thebibliography}
\end{document}